\documentclass[twocolumn,showpacs,preprintnumbers,amsmath,amssymb]{revtex4}

\usepackage{graphicx}% Include figure files
\usepackage{dcolumn}% Align table columns on decimal point
\usepackage{bm}% bold math
\newcommand {\apgt} {\ {\raise-.5ex\hbox{$\buildrel>\over\sim$}}\ }
\newcommand {\aplt} {\ {\raise-.5ex\hbox{$\buildrel<\over\sim$}}\ }

\begin{document}

\title{Universal nonmonotonic structure in the 
saturation curves of MOT-loaded Na$^+$ ions 
stored in an ion-neutral  hybrid trap: Prediction and observation} 

\author{R. Bl\"umel$^1$, J. E. Wells$^2$, D. S. Goodman$^{2,3}$, J. M. Kwolek$^2$, 
and W. W. Smith$^2$} 
\affiliation{$^1$Department of Physics, Wesleyan University, Middletown, 
Connecticut 06459, USA} 
\affiliation{$^2$Department of Physics, University of Connecticut, Storrs, 
Connecticut 06269, USA} 
\affiliation{$^3$Department of Sciences, Wentworth Institute of Technology, 
Boston, Massachusetts 02115, USA}

%\affiliation{Department of Physics, Wesleyan University, 
%Middletown, Connecticut 06459-0155}
  
\date{\today}

\begin{abstract} 
We predict that 
the maximal, steady-state ion capacity 
$N_s(\lambda)$ of radio-frequency (rf) traps, 
loaded at a rate of 
$\lambda$ particles per rf cycle, shows universal, 
nonlinear, nonmonotonic 
behavior as a function of loading rate $\lambda$. 
The shape of $N_s(\lambda)$, characterized by 
four dynamical regimes, 
is universal, i.e., it is predicted to 
manifest itself in {\it all types} of rf traps independent 
of the details of their construction and independent of 
particle species loaded. 
For $\lambda\ll$ 1 (Region I), as expected, 
$N_s(\lambda)$ increases monotonically 
with $\lambda$. However, contrary to intuition, at intermediate 
$\lambda \sim 1$  (Region II), 
$N_s(\lambda)$ reaches a maximum, 
followed by a local 
minimum of $N_s(\lambda)$ (Region III).
For $\lambda\gg 1$ (Region IV), $N_s(\lambda)$ 
again rises monotonically. In Region IV 
numerical simulations, analytical calculations, and experiments show 
$N_s(\lambda)\sim \lambda^{2/3}$. 
We confirm our predictions both experimentally with MOT-loaded 
Na$^+$ ions stored in a 
hybrid ion-neutral trap and numerically with the help of 
detailed ab-initio molecular-dynamics simulations. 
\end{abstract}

\pacs{37.10.Ty,     % ion trapping 
           52.27.Jt,      % Nonneutral Plasmas
           52.50.Qt}     % Plasma heating by rf fields 


\maketitle

%
%=================================================================================
%
%
\section{Introduction}
\label{INTRO}
Radio-frequency (rf) traps \cite{Paul1,PKG} are important devices in 
widespread use for the long-time storage of charged particles. 
These traps come in a multitude of shapes  and 
sizes \cite{Paul1,PKG,shs1,shs2,shs3,shs4} 
and their applications range from high-resolution 
spectroscopy \cite{HRS} to atomic clocks 
\cite{WC1} 
%\cite{WC1,WC2} 
and quantum computers
\cite{QCOMP}. They have also been used 
in nonlinear dynamics for the investigation of nonlinear 
phenomena ranging from crystallization 
\cite{MPQcryst,Wcryst,Nature,Kappler}
to the 
investigation of strange attractors \cite{BH}.  
A fundamental problem of great theoretical and practical interest 
is the maximal ion capacity $N_s(\lambda)$ of rf traps loaded at a 
constant rate of 
$\lambda$ particles per rf cycle. 
While the absolute value of $N_s$ 
depends on the trap's physical size 
and the details of its construction, 
the qualitative dependence of
$N_s(\lambda)$ on $\lambda$ does not. 
In fact, we found that the shape of $N_s(\lambda)$ 
is universal, 
i.e., it is the same for any kind of rf trap, 
and shows four clearly defined dynamical 
regimes, which we label Regions I to IV. 
In this paper, based on physical 
arguments and detailed ab-inito molecular-dynamics 
simulations, we predict 
the qualitative shape of the universal curve $N_s(\lambda)$, 
and experimentally verify our 
predictions 
with the help of a MOT-loaded ion-neutral 
trap \cite{UCT}. While 
knowledge of the steady-state ion capacity 
$N_s$ of rf traps in general is in itself an 
important fundamental problem, the results are 
also of practical interest in atomic physics. 
Collision-rate experiments 
\cite{UCT,Grier,Sulli,RLSWR,Rangwala,RJRR}, e.g., use the 
steady-state ion capacity to measure the total collision rate,  
because it ensures constant density, 
size, and temperature conditions during the measurement. This is 
particularly helpful when working with optically 
dark, closed-shell ions such as 
Rb$^+$ and Na$^+$. 
 
Our paper is organized as follows. In Sec.~\ref{THEO} we 
present the basic dynamical equations for the three-dimensional 
(3D) 
Paul trap and the linear Paul trap used in our 
molecular dynamics simulations together with the methodology 
according to which our simulations are performed. In this 
section we also present the theoretical evidence for the 
four nonlinear loading regimes encountered in these two 
trap types. In Sec.~\ref{EXP} we present our experimental 
evidence that confirms the prediction 
of the four different 
dynamical regimes. In particular, in the case of the linear 
Paul trap, 
we confirm the 
presence of the dip 
characterizing the dynamical Region III. 
In Sec.~\ref{DISC} we discuss our results. We summarize 
and conclude our paper in Sec.~\ref{SC}. We also provide 
an appendix in which we derive the differential equation 
whose stationary solution yields 
the fundamental scaling relation  
$N_s(\lambda)\sim \lambda^{2/3}$ in Region IV. 
%

%==========================================
 
\section{Theory}
\label{THEO}
Initially, we discovered the non-linear, 
nonmonotonic structure of 
$N_s(\lambda)$ 
in molecular-dynamics loading simulations 
of a 3D 
Paul trap. 
Denoting by $r_0$ and $z_0$ the distances of the 
ring electrode and the end-cap electrodes from the trap's 
center, respectively, by $U_0$ and $V_0$ the amplitudes of the 
dc and ac voltages applied to the trap, respectively, 
by $\Gamma$ the damping constant 
(generated, e.g., by laser cooling \cite{Nature}), 
and by $\omega$ the (angular) frequency of 
the trap's ac voltage, 
the (dimensionless) equations of motion 
of $N$ particles in the trap [$\vec r=(x,y,z)$] are  \cite{TNB,MF} 
\begin{equation} 
\ddot{\vec r}_i + \gamma \dot{\vec r}_i + 
[a-2q\sin(2 t)]
 \left( \begin{matrix} x_i \\ y_i \\ -2 z_i \end{matrix} \right) 
 =   \sum_{\substack{j=1\\j\neq i}}^N 
 \frac{\vec r_i - \vec r_j}{|\vec r_i - \vec r_j |^3}, 
\label{Eq1} 
\end{equation} 
where $i=1,\ldots,N$ labels the trapped particles, 
\begin{equation}
q = \frac{4Q V_0}{m\omega^2(r_0^2+2z_0^2)},\ \ \ 
a = \left( \frac{2U_0}{V_0}\right) q, 
\label{Eq2}
\end{equation}
are the two control parameters of the Paul trap 
\cite{Paul1,Nature,Kappler}, 
time is measured in units of 
\begin{equation}
\tau_0=\frac{2}{\omega}, 
\label{tau0}
\end{equation}
distances are measured in units of 
\begin{equation}
l_0=(Q^2/\pi\epsilon_0 m \omega^2)^{1/3}, 
\label{l0}
\end{equation}
where $Q$ is the charge and $m$ is the 
mass of each of the trapped particles, 
$\epsilon_0$ is the permittivity of the vacuum, 
and 
\begin{equation}
\gamma=\tau_0 \Gamma = \frac{2\Gamma}{\omega}
\label{gdef}
\end{equation}
is the dimensionless damping constant. 
We use dimensionless quantities in this section 
because only this way is it possible to see that 
the equations of motion of particles in the Paul trap 
do not depend on the six physical parameters 
$r_0$, $z_0$, $\Gamma$, $\omega$, $Q$, and $m$ separately, 
but only on the three scaled, dimensionless parameters $a$, $q$, 
and $\gamma$. Therefore, instead of the need to explore a
six-dimensional parameter space, which is practically impossible, 
we only need to explore a three-dimensional parameter space. 
This grows to four dimensions, if we include the loading rate 
$\lambda$. 
 
In our simulations particles are 
created, one at a time, at times $t_k$, $k=1,2,\ldots$, 
either with zero initial velocity (a good approximation 
for MOT-loaded ions \cite{SIMION}) or with a 
thermal velocity distribution as discussed below. 
Assuming that the creation times are uncorrelated,  
the time intervals $\Delta t_k=t_{k+1}-t_k$ are Poissonian 
distributed with probability distribution 
$P(\Delta t)=\lambda \exp(-\lambda \Delta t)$. 
Concerning their spatial distribution, we assume 
that the particles are created at random positions 
with uniform distribution 
within an ellipsoidal volume with semi-major axes 
$L_x,L_y,L_z$, centered 
at the origin of the trap. This includes the case of a 
spherical loading zone of radius $R$, in which case 
we have $R=L_x=L_y=L_z$. 
In the time interval between 
any two creation events, i.e., for $t_k < t < t_{k+1}$,
the particles in the 
trap are governed by the equations of motion 
(\ref{Eq1}). 
When arriving at $t_{k+1}$, and before creating the next 
particle, we check whether one or more particles 
have left the trap by crossing an absorbing boundary. 
The existence of an absorbing boundary is a fundamental 
property of all rf traps, which determines and 
limits the storage capacity of any given trap. 
This boundary may be due to any number of 
unavoidable physical causes, such as 
the trap's electrodes, or 
instabilities induced by higher-order rf multipoles \cite{shs3}. 
Since all traps are constructed differently, and to show that 
our predicted effect is robust with respect to various 
geometries of absorbing boundaries, we used spherical 
boundaries of radius $R_{\rm sph}$, boxes with side lengths 
$2x_{\rm box}, 2y_{\rm box},2z_{\rm box}$, 
and cylinders with radius $R_{\rm cyl}$ 
in the $x$-$y$ plane and length 
$2z_{\rm cyl}=\{[q^2/(4B)]-1/2\}^{1/2}$ in $z$ direction to 
cover a wide variety of possible boundary geometries. 
Following instantaneous deletion 
of all particles that exceed the confines of the 
absorbing boundary,
the next particle is loaded at 
$t=t_{k+1}$. This procedure is followed for all $t_k$ until a 
pre-specified maximal simulation time is reached. 
%
%-----------------------------------------------------------------------
\begin{figure}
\centering
\includegraphics[scale=0.55,angle=0]{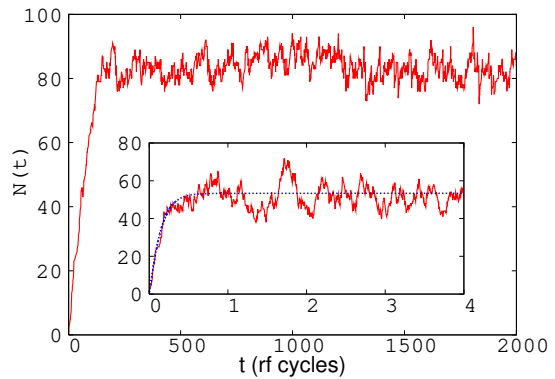}
\caption{\label{fig1} (Color online) 
Number of 
particles $N(t)$ in a 3D Paul trap 
as a function of $t$ (in rf cycles), 
loaded with loading rate $\lambda=1$ particle / rf cycle  
inside of a spherical volume of radius $R=3$ centered 
at the origin of the trap. Particles are absorbed at a critical 
distance $R_c=15$. Trap control parameters are $\gamma=0$, 
$q=0.2$, $a=0.02$.  Inset: Test of the loading equation (\ref{NDE}) 
for a loading rate $\lambda=300$ particles / rf cycle. Red solid line: 
Number of particles $N(t)$ obtained via numerical simulation of 
the loading process. Blue dotted line: Prediction 
according to  (\ref{EX}). 
       }
\end{figure}
%--------------------------------------------------------------------------
%
 
The result of a typical trap loading 
simulation for a spherical loading zone with $R=3$ 
is shown in Fig.~\ref{fig1}. 
The red, fluctuating line in 
Fig.~\ref{fig1} shows the time evolution of the 
particle number $N(t)$ in the trap for $\lambda=1$ as a function of $t$ 
(in rf cycles) for $q=0.2$, $a=0.02$, 
and $R_{\rm sph}=15$. 
A near linear rise of $N(t)$ is followed by a sharp bend into a steady state 
in which $N(t)$ fluctuates around $N_s=\langle N(t)\rangle_t$, the 
time average of $N(t)$ in the steady-state. 
 
%
%-----------------------------------------------------------------------
\begin{figure}
\centering
\includegraphics[scale=1.1,angle=0]{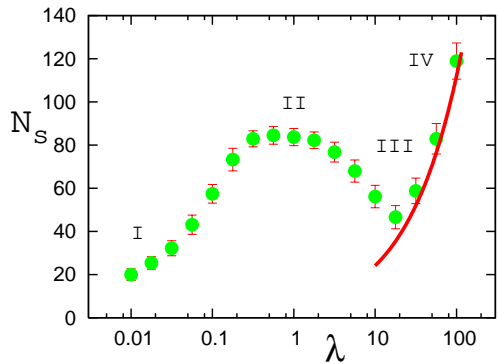}
\caption{\label{fig2} (Color online) 
Nonmonotonic loading curve $N_s(\lambda)$ for a 3D Paul trap 
with $\gamma=0$, 
$q=0.2$, $a=0.02$, $R=3$, and $R_{\rm sph}=15$. Solid green dots: 
Results of 3D molecular dynamics loading simulations. The red bars 
indicate the amplitudes of the $N(t)$ fluctuations in the saturated state. 
For the first two dots the $N(t)$ fluctuations are smaller than the 
plot symbols. 
The four distinct regions of the loading curve are labeled 
I to IV. 
The red solid line is the curve 
$N_s(\lambda)=5.2\times \lambda^{2/3}$, 
which confirms the $\sim \lambda^{2/3}$ behavior of 
3D Paul traps in Region IV.
       }
\end{figure}
%--------------------------------------------------------------------------
%

%
%-----------------------------------------------------------------------
\begin{figure}
\centering
\includegraphics[scale=1.1,angle=0]{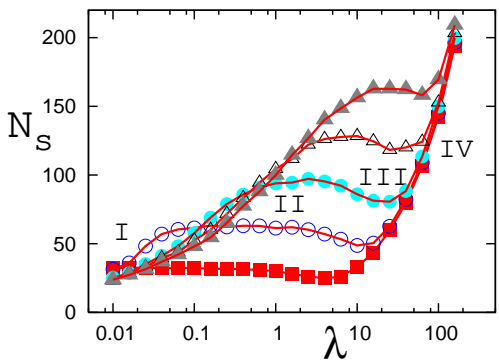}
\caption{\label{fig3} (Color online) 3D Paul trap capacities 
$N_s(\lambda)$ for various trap control parameters $q$ 
and $\gamma=0$. 
Solid red squares:  $q=0.1$; 
open blue circles:  $q=0.15$; 
solid cyan circles: $q=0.2$; 
open black triangles: $q=0.25$;
solid grey triangles: $q=0.3$. The trap parameter 
$a$ associated with each of the five $q$ values 
is $a=q^2/2$. 
The four dynamical regimes are labeled I to IV. 
The solid red lines connecting the data points are drawn 
to guide the eye. All five curves asymptote to 
$N_s(\lambda)\sim \lambda^{2/3}$ for large $\lambda$. 
       }
\end{figure}
%--------------------------------------------------------------------------
%
For many applications (see, e.g., \cite{UCT,Rangwala}) it is necessary to know 
$N_s(\lambda)$ for a given rf trap. Figure~\ref{fig2} shows $N_s(\lambda)$ 
as a function of loading rate $\lambda$ for the same 
Paul trap control parameters 
used to generate Fig.~\ref{fig1}. 
The resulting curve in Fig.~\ref{fig2} 
clearly shows four distinct regimes, labeled 
I to IV. 
Region IV is the most straightforward to 
understand physically; therefore, we discuss it first. In Region IV, the loading 
rate is so large that a large space-charge density develops in the loading 
region. The resulting large electric field 
accelerates the 
loaded particles outward toward the absorbing boundary
where the particles are lost from the trap. 
Since in this case the forces due to the space charge 
completely overwhelm the forces due to the trap fields, 
the particles' dynamics are accurately described 
as a Coulomb explosion \cite{CE}. Denoting 
the number of particles inside of the loading zone 
by $\tilde N(t)$ [in contrast to $N(t)$, which refers to 
the total number of particles in the trap], the 
temporal evolution of the number of particles, $\tilde N(t)$, 
inside of the loading zone 
is governed by the differential equation (see appendix): 
\begin{equation}
\frac{d\tilde N(t)}{dt} = \tilde\lambda - \left( \frac{ \tilde N(t)}{\tilde R}\right)^{3/2},  
\label{NDE}
\end{equation}
where $\tilde R = [16/(9\pi)]^{2/3} R\approx R$ is the effective 
radius of the loading volume and $\tilde\lambda$ is the 
number of particles loaded per unit of  
dimensionless time. According to (\ref{lamtil}), $\tilde\lambda$ is 
related to $\lambda$, the number of particles loaded per 
rf cycle by $\tilde\lambda=\lambda/\pi$. 
In the stationary state, we have $d\tilde N(t)/dt=0$. 
Therefore, we obtain from (\ref{NDE}) 
\begin{equation}
\tilde N_s(\lambda) = \tilde R\, \tilde\lambda^{2/3} = 
\left(\frac{16}{9\pi^2}\right)^{2/3} R\  \lambda^{2/3} . 
\label{NSS}
\end{equation}
Because of continuity, the steady-state number of particles in 
the trap, $N_s(\lambda)$, is proportional to the number of 
particles in the loading region, $\tilde N_s(\lambda)$. 
Therefore, 
the $\lambda^{2/3}$ dependence of $\tilde N_s(\lambda)$ is 
reflected in Fig.~\ref{fig2} (solid red line). 
Because there are more particles in the trap than there are in the 
loading zone, the pre-factor 5.2 of $\lambda^{2/3}$, stated  
in the caption of Fig.~\ref{fig2}, is larger than 
the pre-factor $(16/\pi^2\sqrt{3})^{2/3}\approx 1$ of $\lambda^{2/3}$, 
computed from (\ref{NSS}) with $R=3$.  
 
The solution of (\ref{NDE}) can be stated implicitly in closed form: 
\begin{align} 
t &= -\frac{2\tilde R}{3\tilde\lambda^{1/3}} \Big\{ \ln(\alpha-\tilde N^{1/2}) - \frac{1}{2} 
\ln(\tilde N+\alpha \tilde N^{1/2} + \alpha^2)  \cr
 &+ \sqrt{3} \arctan\left( \frac{2 \tilde N^{1/2} + \alpha}{\alpha\sqrt{3}} \right)  
      -\sqrt{3} \arctan\left( \frac{1}{\sqrt{3}} \right)   \Big\} , 
\label{EX}
\end{align}
where $\alpha=\tilde R^{1/2} \tilde\lambda^{1/3}$. That (\ref{EX}) is indeed 
a solution of (\ref{NDE}) may be checked immediately by 
differentiating (\ref{EX}) with respect to $t$. 
Since the trap potentials are not important in Region IV, the results 
(\ref{NDE}) -- (\ref{EX}) apply universally to {\it all} rf traps, 
for instance 3D or linear Paul traps.  

In order to test (\ref{EX}), we chose $\tilde R=3$ and a large 
loading rate of $\lambda=300$ 
particles / rf cycle to be sure that we are in Region IV. 
The result of the corresponding 
molecular dynamics simulation 
of $N(t)$ is shown as the solid, fluctuating red line in the inset of Fig.~\ref{fig1}. 
The blue, dotted line in the inset of Fig.~\ref{fig1} is the 
prediction according to (\ref{EX}). Both agree perfectly within 
the expected fluctuations of $N(t)$, which are due to the 
Poissonian loading process. 
  
While for large $\lambda$ the trap potentials are not important, 
they become progressively more important when $\lambda$ is lowered. 
In this case, for low enough $\lambda$, 
particles created close to the edge of the loading zone no longer 
have enough energy to overcome the trap potentials 
and are reflected back into the interior of the trap. 
When 
back-reflection occurs, fewer particles escape and 
$N_s(\lambda)$ increases for decreasing $\lambda$ up to 
a maximum in Region II, effectively creating a dip 
in Region III (see Fig.~\ref{fig2}). 
However, the reflected particles will not stay in the trap forever. 
Due to rf heating \cite{Nature,TNB}, and given enough time 
(small loading rates), these particles will heat out of the trap, eventually 
lowering the number $N_s(\lambda)$ of stationary-state particles 
in the trap below the bottom of the dip in Region III (see Fig.~\ref{fig2}), 
thus explaining both the formation of the 
maximum in $N_s(\lambda)$ 
(Region II in Fig.~\ref{fig2}) and 
the eventual 
decline of $N_s(\lambda)$ in the direction of ever diminishing loading rates
(Region I in Fig.~\ref{fig2}). Since rf heating is a universal feature 
of all rf charged-particle traps, we predict that the qualitative 
shape of $N_s(\lambda)$ is 
universal for all rf traps. 
 
%
%-----------------------------------------------------------------------
\begin{figure}
\centering
\includegraphics[scale=1.2,angle=0]{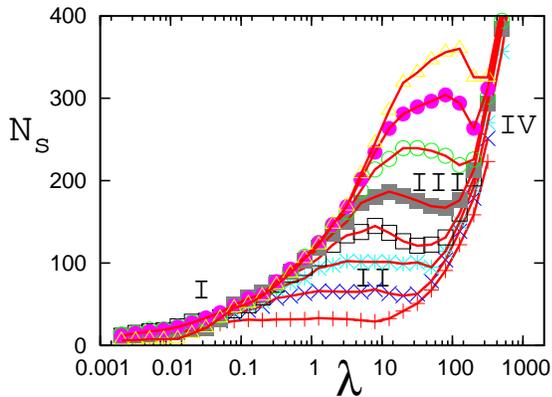}
\caption{\label{fig4} (Color online) 
Saturation curves $N_s(\lambda)$, for a linear Paul trap. 
Red pluses: $q=0.35$; 
blue crosses $q=0.40$;
cyan diamonds: $q=0.45$; 
black open squares:  $q=0.50$; 
grey solid squares:  $q=0.55$; 
green open circles: $q=0.60$; 
magenta solid circles: $q=0.65$;
brown open triangles: $q=0.70$. 
The solid red lines connecting the data points are drawn 
to guide the eye. All eight curves asymptote to 
$N_s(\lambda)\sim \lambda^{2/3}$ for large $\lambda$. 
The labels I to IV refer to the four different dynamical regimes. 
       }
\end{figure}
%--------------------------------------------------------------------------
 
To strengthen the claim of universality of the nonmonotonic 
curve shown in Fig.~\ref{fig2}, we show in Fig.~\ref{fig3} 
the saturated number of particles 
$N_s(\lambda)$ 
as a function of loading rate $\lambda$ for several 
values of $q$ with $a=q^2/2$, $L_x=L_y=3$, $L_z=6$, 
and $x_{\rm box}=y_{\rm box}=z_{\rm box}=15$. 
All five curves in Fig.~\ref{fig3} 
clearly show all four dynamical regimes, again labeled 
I to IV. This shows that the nonmonotonic behavior is 
robust with respect to 
(a) a change in $q$, 
(b) a change in the geometry of the loading zone 
      (spherical in Fig.~\ref{fig2}; elliptical in Fig.~\ref{fig3}), and 
(c) a change in 
the geometry of the absorbing boundary (spherical in 
Fig.~\ref{fig2}; cubic in Fig.~\ref{fig3}). This applies 
in particular to Region IV, in which 
all five curves in 
Fig.~\ref{fig3} are seen to converge to 
the same $\sim \lambda^{2/3}$ asymptote. 
The independence of $a$ and $q$ in region IV 
is explained by (\ref{NDE}), which is independent 
of the trap potentials and therefore independent 
of the trap parameters $a$ and $q$. Figure~\ref{fig3} 
also shows that the dip becomes shallower 
with both increasing and decreasing $q$ and shifts to the right with 
increasing $q$. In addition,  we see that Region II becomes more extended with 
decreasing $q$. At present we do not have a theoretical 
explanation for these observed effects. 
 
To check the universality of the nonmonotonic behavior, 
we also simulated a linear Paul trap 
\cite{LPT} whose equations of motion, in the notation and units 
of (\ref{Eq1}), 
are given by 
\begin{equation}
\left(   \begin{matrix}   
                 \ddot x_i + [a-2q\sin(2t)] x_i - Bx_i \cr
                \ddot y_i - [a-2q\sin(2t)] y_i  - By_i \cr
                \ddot z_i + 2Bz_i  \cr 
          \end{matrix} \right) 
   = \sum_{\substack{j=1 \\ j\neq i}}^{N_k} 
   \frac{   \vec r_i - \vec r_j} {|\vec r_i - \vec r_j|^3 }, 
\label{LPT}
\end{equation}
where $B$ is a positive constant. 
We simulate the loading 
process of the linear Paul trap 
in analogy to the 3D Paul trap  
as discussed above with a loading region of 
radius $R=3$, 
located at the geometric center of the linear trap and 
a cylindrical absorbing boundary with 
$R_{\rm cyl}=15$. 
This time, however, we 
use the equations of motion (\ref{LPT}) between creation times.  
Figure~\ref{fig3} shows the resulting $N_s(\lambda)$ for 
eight different $q$ values with $a=0$, and $B=0.042$. 
We clearly see the 
four different regions of $N_s(\lambda)$, observed previously 
in the case of the 3D Paul trap. This 
provides 
corroborating evidence for the 
universal nature of the shape of $N_s(\lambda)$ for 
all types of rf traps. 

%
%-----------------------------------------------------------------------
\begin{figure}
\centering
\includegraphics[scale=0.55,angle=-0]{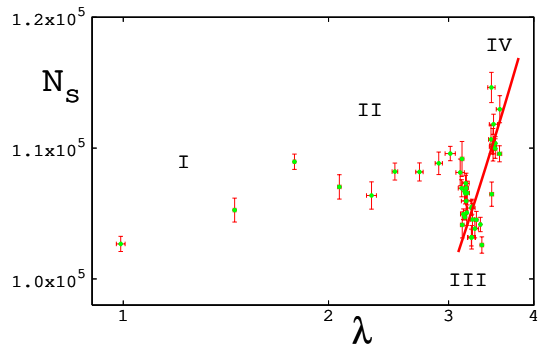}
\caption{\label{fig5} (Color online) Experimental saturation curve for the 
hybrid trap \cite{UCT}, 
operated with $q=0.26$. All four predicted dynamical regimes 
are present. 
% Region I:    $\lambda<85\,{\rm V/s}$; 
% Region II:   $85\,{\rm V/s} < \lambda < 107\,{\rm V/s}$; 
% Region III:  $107\,{\rm V/s} < \lambda < 118\,{\rm V/s}$; 
% Region IV: $\lambda > 118\,{\rm V/s}$. 
The red solid line 
is the curve $N_s=48000 \lambda^{2/3}$. 
The error bars represent the statistical 
errors associated with multiple data runs. 
The zero-point of the horizontal scale is suppressed. 
       }
\end{figure}
%--------------------------------------------------------------------------
%

%==========================================
\section{Experiment}
\label{EXP}
The ultimate test of our 
theoretical predictions is an experiment. For this purpose we
used a MOT-loaded ion-neutral hybrid trap \cite{UCT}. 
Figure~\ref{fig5} shows the results of our experiments. 
All four predicted dynamical regimes are present and 
the shape of the experimental $N_s(\lambda)$ curve is seen to be 
qualitatively the same as predicted by the model simulations and 
our qualitative analysis of the physical mechanisms that determine 
the steady-state populations in rf traps. From the experimental data 
we obtain $\epsilon=0.745\pm 0.098$ 
for the exponent $\epsilon$ 
in $N_s(\lambda)\sim \lambda^{\epsilon}$, 
which 
is consistent with the predicted value $\epsilon=2/3$ 
in Region IV (red solid line in Fig.~\ref{fig5}). 
We note that, corroborating the universality claim, the qualitative shape 
of the nonlinear, nonmonotonic behavior of $N_s(\lambda)$ 
was observed in all of 
our simulations and experiments, 
independent of rf amplitudes, rf frequencies, and MOT 
sizes. With a modulation depth of about 
5\%, the dip in Fig.~\ref{fig5}
is small. However, noticing that the experimental $q$ of 
$q=0.26$ is small, this observation is consistent 
with Fig.~\ref{fig4}, 
which shows that the modulation depth of the dip decreases with 
decreasing $q$. 

%=========================================
% Discussion
%=========================================
%Whenever we extracted $N_s(\lambda)$ from the simulation data, we made 
%sure that we simulated the steady state 
%over a sufficiently large number of rf cycles to 
%be able to extract $N_s$ reliably. For the first few data points 
%in Figs.~\ref{fig2} and \ref{fig3}, e.g., this required up to 500,000 
%rf cycles of simulation time. 
%
%
 
%==========================================
 
\section{Discussion}
\label{DISC}
To strengthen the universality claim and to emphasize 
the robustness of our predictions, 
we performed numerous additional simulations that all  
confirmed that the qualitative shape of the saturation curve, 
characterized by its four dynamical regimes, is insensitive to both 
the type of traps used and their particular loading mechanisms. 
In particular, we performed 
additional simulations 
with (a) various shapes of the loading ellipse with 
aspect ratios of $L_x:L_y:L_z$ up to $1:1:4$, (b) replacing the Poissonian 
distribution of loading times with a uniform distribution, 
(c) replacing the uniform 
spatial distribution with a Gaussian distribution, (d) various 
geometries of the absorbing boundary and boundary locations, 
and (e) increasing the number of trapped particles to up to 1000 
by changing the diameter of the absorbing boundary. 
All five numerical 
tests confirmed the qualitative shape 
of $N_s(\lambda)$ as shown in Figs.~\ref{fig2} -- \ref{fig5}. 
In addition, we performed the following two checks 
concerning ion cooling and the effect of temperature. 
 
In many ion-trap experiments, strong 
laser cooling is continuously switched on, even during 
the loading stage (see, e.g., \cite{Nature}). 
Therefore, to test the influence of cooling during 
trap loading, we ran additional 3D Paul-trap loading 
simulations with strong damping switched on,
so strong in fact, that it would crystallize the ions
\cite{MPQcryst,Wcryst,Nature,LPT} 
if we were not constantly loading. 
This corresponds to $\gamma$ values between 
$5\times 10^{-4}$ and $10^{-3}$. 
As a result we find that we still have all four 
dynamical regimes, in particular the dip.
Moreover, the $N_s(\lambda)$ curves in the two cases
(with and without damping, respectively) are nearly identical, 
differing from each other only within the natural fluctuations 
of $N_s$ due to the Poissonian loading process. 
In some of our more recent simulations, 
just to push the envelope,   
we increased the damping to five times the one 
needed for crystallization ($\gamma=5\times 10^{-3}$). 
It still did not have any 
effect on the qualitative shape of the loading curves. 
Thus, we conclude that the presence of laser cooling that can 
realistically be achieved experimentally has no effect on 
our predicted phenomena. We also mention that 
a large laser cooling power does not necessarily 
result in a large $\gamma$, since $\gamma$ 
represents the balance between laser cooling and 
the substantial amount of heating caused by 
the loading process. 
 
In our experiments we load from a cold MOT and 
the assumption of zero kinetic energy at the 
instance of charged-particle creation is justified. 
However, in case the trap is loaded by ionizing 
the rest gas or via a thermal neutral beam, the 
particles may be created with substantial 
initial kinetic energy. To test the effect of initial 
kinetic energy on our predictions, we performed 
additional loading simulations, 
imparting a random velocity on each particle 
at the instance of its creation whose energy 
equivalent was up to 20\% of the trap depth. 
In our experiments this would be equivalent to 
about room temperature. 
Even in this case we are 
still able to observe all four dynamical 
regions. The explanation is that in steady-state 
there is an equilibrium between particles created 
and particles leaving the trap. Those that leave, 
already have kinetic energies of the order of 
the trap's depth, which they are able to impart, 
via Coulomb collisions, to newly created particles. 
Therefore, the fact that particles are created with 
a kinetic energy less than the 
trap's depth is only a minor perturbation on the 
energetic particle dynamics that is already 
going on inside of the trap. 
 
In our simulations we found that the dip in Region~III is 
more pronounced if the absorbing boundaries are 
further from the edges of the loading zone. This observation explains 
why the dip is more pronounced in Fig.~\ref{fig2} compared 
with the dips in Fig.~\ref{fig3}. In Fig.~\ref{fig3} we used 
an elliptical loading zone with $L_z=6$ compared with 
the spherical loading zone in Fig.~\ref{fig2} with $R=3$. 
Therefore, in $z$ direction, it is easier for the particles 
in Fig.~\ref{fig3} to bridge the gap to the absorbing 
boundary, because many of them are already created 
closer to the absorbing boundary. The shorter distance 
to cover results in a shallower dip.

%====================================================
 
\section{Summary and Conclusions}
\label{SC}
In this paper we report the discovery of 
the non-monotonic shape  
of the saturated ion number $N_s(\lambda)$ 
of rf traps as a function of 
loading rate $\lambda$ and 
present evidence for its universality. 
Four dynamical regions are predicted. In Region I, 
$N_s(\lambda)$ increases monotonically with $\lambda$, 
reaching a maximum (Region II) at intermediate loading rates 
$\lambda$, followed by a valley (Region III), and an ultimate 
$\sim \lambda^{2/3}$ increase of $N_s(\lambda)$ for very large 
loading rates. We argue that the four regions are expected 
on the basis of physical reasons and are caused by the 
interplay between trap potentials, space-charge effects, 
and rf heating. 
The validity of our predictions, in particular their universality, 
are corroborated with 
the help of ab-initio molecular-dynamics simulations of 
a 3D Paul trap and a linear Paul trap, which both show 
the predicted qualitative shape of $N_s(\lambda)$. 
The theoretical predictions are confirmed experimentally 
with the help of Na$^+$ ions in a MOT-loaded linear 
Paul trap. 
 
Apart from our research group \cite{UCT}, several other groups 
\cite{Grier,Sulli,RLSWR,Rangwala,RJRR}
have the necessary experimental facilities to test our predictions 
in the case of MOT-loaded traps, and since, as we showed, 
the predicted phenomenon 
is robust with respect to rf trap types, loading mechanisms, and 
temperature effects, we hope that other research groups may soon 
test and confirm our predictions. 
 
\section{Acknowledgement} 
Financial support by NSF grant number 1307874 is gratefully 
acknowledged. 
 
\appendix*
\section{Region IV differential equation} 
\label{APP} 
In this appendix we present a simple, explicitly solvable, analytical 
model that reproduces the $\lambda^{2/3}$ scaling of $N_{s}$ in 
Region IV, the fast-loading regime. While in Sec.~\ref{THEO} 
we used dimensionless quantities to bring out the scaling 
properties of the Paul-trap equations, it is more convenient 
in this appendix to derive our equations using SI units. 
In these units, we denote time (measured in seconds) by 
$\tau$, the loading rate (measured in particles per second) by $\Lambda$, 
and the radius of the loading zone 
(measured in meters) by $\hat R$. 
However, in order to make contact with the formulas in 
Sec.~\ref{THEO}, it is convenient to define 
$\lambda$, as we did in Sec.~\ref{THEO}, as the 
number of particles loaded per rf cycle, and by 
\begin{equation}
\tilde\lambda = \tau_0 \Lambda = 2\Lambda/\omega = \lambda/\pi 
\label{lamtil}
\end{equation}
the dimensionless loading rate per unit of dimensionless time. 
The unit of time, $\tau_0$, in (\ref{lamtil}) is defined in 
(\ref{tau0}). The last equality in (\ref{lamtil}) comes about since 
in dimensionless time an rf cycle has a length of $\pi$ 
[see (\ref{Eq1})]. 
 
We are now ready to start our derivation of equation (\ref{NDE}) 
in Sec.~\ref{THEO}, which holds in Regime IV. 
As discussed in 
Sec.~\ref{THEO}, in this regime 
we may neglect the trap potential altogether. We assume that 
the trapped particles are created at random locations inside of a 
sphere of radius $R$ 
with uniform probability distribution and 
loading rate $\lambda$. 
If there are ${\tilde N}$ particles present inside 
the sphere of radius $\hat R$, the charge density, approximated 
as a continuous distribution, is 
\begin{equation}
\rho = \frac{3{\tilde N}Q}{4\pi \hat R^3}, 
\label{rhoequ}
\end{equation}
where $Q$ is the charge of each trapped particle. 
Using Gauss' law, the radial electric field, pointing outward, a distance 
$r$ away from the center of the sphere, is 
\begin{equation}
E = \frac{\rho r}{3\epsilon_0}. 
\label{Efield}
\end{equation}
Therefore, the radial, outward directed force experienced by an ion 
a distance $r$ away from the loading sphere is 
\begin{equation}
F = \left( \frac{\rho Q r}{3\epsilon_0}\right) = \left( \frac{{\tilde N}Q^2}{4\pi\epsilon_0 \hat R^3}
\right)\ r. 
\label{force}
\end{equation}
This equation shows that 
the force experienced by a single trapped particle is like the harmonic force of 
an inverted oscillator. Therefore the equation of motion of the ion is 
\begin{equation}
m\ddot r = F\ \ \ \implies\ \ \ \ddot r = {\tilde N} \Omega^2 r,
\label{eqmo}
\end {equation}
where $m$ is its mass and 
\begin{equation}
\Omega = \left( \frac {Q^2}{4\pi\epsilon_0 m \hat R^3}\right)^{1/2}  . 
\label{Ome}
\end{equation}
The general solution of (\ref{eqmo}) is 
\begin{equation}
r(\tau) = A \exp(\sqrt{{\tilde N}}\Omega \tau ) + B \exp(-\sqrt{{\tilde N}}\Omega \tau ),
\label{gsol}
\end{equation}
where $A$ and $B$ are constants. If we assume that at $\tau =0$ 
the particle is created a distance $s$ away from the center 
of the loading sphere, (\ref{gsol}) may be written as 
\begin{equation}
r(\tau) = s \cosh (\sqrt{\tilde N} \Omega \tau), 
\label{rsol}
\end{equation}
and the velocity of the particle is 
\begin{equation}
\dot r(\tau) = s \sqrt{{\tilde N}} \Omega \sinh(\sqrt{{\tilde N}} \Omega \tau). 
\label{vsol}
\end{equation}
Define $T_0$ as the time it takes the particle to reach the rim of 
the loading sphere at radius $\hat R$ if the ion starts at radius 
$s$ with zero velocity. Then: 
\begin{align}
\hat R &= s \cosh(\sqrt{{\tilde N}} \Omega T_0) \\ 
    &\implies \cosh(\sqrt{{\tilde N}} \Omega T_0) = \frac{\hat R}{s}. 
\label{Requ}
\end{align}
When it arrives at $\hat R$, the velocity of the ion is 
\begin{equation}
v_0 = s \sqrt{{\tilde N}} \Omega \sinh(\sqrt{{\tilde N}} \Omega T_0).
\label{v0eq}
\end{equation}
Use (\ref{Requ}), together with $\cosh^2(x)-\sinh^2(x) = 1$,  
to write (\ref{v0eq}) in the form 
\begin{equation}
v_0 = \sqrt{{\tilde N}} \Omega \sqrt{\hat R^2 - s^2}.
\label{v0expl}
\end{equation}
Next, we compute the average velocity $\bar v_0$ with which 
a randomly created particle arrives at $\hat R$. 
Denoting by $\cal{V}$ the volume of the loading sphere, 
${\cal V}=4\pi \hat R^3/3$, and using the fact that we assume a
uniform probability distribution of particle creation positions 
within the sphere, we obtain: 
\begin{align}
\bar v_0 &= \frac{\sqrt{{\tilde N}}\Omega}{\cal{V}} 
\int_{\cal{V}} \sqrt{\hat R^2-s^2}\, dV 
\nonumber \\ 
&= \frac{3\sqrt{{\tilde N}}\Omega}{\hat R^3} 
\int_0^{\hat R} s^2 \sqrt{\hat R^2-s^2}\, ds  
&= \frac{3}{16} \pi \hat R \Omega \sqrt{{\tilde N}}.
\label{v0av}
\end{align}
Since the average ion arrives at $\hat R$ with an average 
velocity $\bar v_0$, directed radially outward, the average 
number $dN_l$ of particles lost from the loading sphere 
in time $d\tau$ is the number of particles in a shell of 
radius $\hat R$ and width $\bar v_0 d\tau$. Explicitly: 
\begin{equation}
dN_l = \frac{\rho}{Q} 4\pi \hat R^2 \bar v_0 d\tau = 
\left( \frac{9\pi}{16}\right) {\tilde N}^{3/2} 
\Omega d\tau.
\label{dNleq}
\end{equation}
The number of particles gained due to loading 
with rate $\Lambda$ is 
\begin{equation}
dN_g = \Lambda d\tau.
\label{dNgeq}
\end{equation}
Therefore the total change $d{\tilde N}$ in the number of particles present 
in the loading sphere is 
\begin{equation}
d{\tilde N} = dN_g - dN_l = \Lambda d\tau - C {\tilde N}^{3/2} d\tau, 
\label{dNeq}
\end{equation}
where 
\begin{equation}
C = \frac{9\pi}{16} \left( \frac{Q^2}{4\pi\epsilon_0 m \hat R^3} \right)^{1/2}. 
\label{Ceq}
\end{equation}
Dividing (\ref{dNeq}) by $d\tau$, we obtain a first-order differential equation 
for the number of particles ${\tilde N}$ inside the loading sphere: 
\begin{equation}
\frac{d{\tilde N}}{d\tau} = \Lambda - C {\tilde N}^{3/2} . 
\label{diffequ}
\end{equation}
To transform this equation into its dimensionless form, 
we use the unit of time, $\tau_0$, and the unit of length, 
$l_0$, defined in (\ref{tau0}) and (\ref{l0}), respectively, 
which relates $\tau$, $\Lambda$, and $\hat R$ in SI units 
to their dimensionless counterparts, $t$, $\lambda$, 
and $R$, respectively, according to [see also (\ref{lamtil}) 
\begin{equation} 
\tau = t \tau_0, \ \ \ \Lambda=\tilde\lambda/\tau_0,\ \ \ 
\hat R = R l_0. 
\label{dimless}
\end{equation}
Using (\ref{dimless}) in (\ref{diffequ}) 
and defining 
\begin{equation}
\tilde R = \left( \frac{16}{9\pi}\right)^{2/3} R, 
\label{RR}
\end{equation}
we arrive at 
\begin{equation}
\frac{d{\tilde N}}{dt} = \tilde\lambda - \left( \frac{\tilde N(t)}{\tilde R}\right)^{3/2}, 
\label{dldeq}
\end{equation}
which is identical with 
(\ref{NDE}) of Sec.~\ref{THEO}. 
% 
  
%===============================================================

\end{document}